\documentclass[conference]{IEEEtran}
\usepackage{fancyhdr}
\IEEEoverridecommandlockouts{}

\usepackage{cite}
\usepackage{amsmath,amssymb,amsfonts}
\usepackage{algorithmic}
\usepackage{graphicx}
\usepackage{textcomp}
\usepackage[table]{xcolor}
\usepackage{xcolor}
\usepackage{easyReview}
\usepackage{verbatim}
\usepackage{booktabs}
\usepackage{multirow}
\usepackage{array}
\usepackage{makecell}
\usepackage{tabularx}
\usepackage{booktabs}
\usepackage{siunitx}

\def\BibTeX{{\rm B\kern-.05em{\sc i\kern-.025em b}\kern-.08em
    T\kern-.1667em\lower.7ex\hbox{E}\kern-.125emX}}

\begin{document}

\title{
\vspace{-0.3cm}
Breaking the Million-Electron and 1 EFLOP/s Barriers: Biomolecular-Scale \emph{Ab Initio} Molecular Dynamics Using MP2 Potentials
\vspace{-0.3cm}
}

\author{
\IEEEauthorblockN{Ryan Stocks}
\IEEEauthorblockA{\textit{School of Computing} \\
\textit{Australian National University}\\
ryan.stocks@anu.edu.au
}
\and

\IEEEauthorblockN{Jorge L. Galvez Vallejo}
\IEEEauthorblockA{\textit{School of Computing} \\
\textit{Australian National University}\\
jorge.galvezvallejo@anu.edu.au
}
\and

\IEEEauthorblockN{Fiona C. Y. Yu}
\IEEEauthorblockA{\textit{School of Computing} \\
\textit{Australian National University}\\
fiona.yu@anu.edu.au
}
\and
\IEEEauthorblockN{Calum Snowdon}
\IEEEauthorblockA{\textit{School of Computing} \\
\textit{Australian National University}\\
calum.snowdon@anu.edu.au
}
\and

\IEEEauthorblockN{Elise Palethorpe}
\IEEEauthorblockA{\textit{School of Computing} \\
\textit{Australian National University}\\
elise.palethorpe@anu.edu.au
\vspace{-0.95cm}
}
\and

\IEEEauthorblockN{Jakub Kurzak}
\IEEEauthorblockA{\textit{Advanced Micro Devices, Inc.}\\
jakub.kurzak@amd.com
\vspace{-0.95cm}
}
\and

\IEEEauthorblockN{Dmytro Bykov}
\IEEEauthorblockA{\textit{Computing for Chemistry} \\
\textit{Oak Ridge National Lab}\\
bykovd@ornl.gov
\vspace{-0.95cm}
}
\and

\IEEEauthorblockN{Giuseppe M. J. Barca}
\IEEEauthorblockA{\textit{School of Computing} \\
\textit{University of Melbourne}\\
giuseppe.barca@unimelb.edu.au
\vspace{-0.95cm}
}

}

\maketitle


\begin{abstract}

The accurate simulation of complex biochemical phenomena has historically been hampered by the computational requirements of high-fidelity molecular-modeling techniques. Quantum mechanical methods, such as \emph{ab initio} wave-function (WF) theory, deliver the desired accuracy, but have impractical scaling for modeling biosystems with thousands of atoms. Combining molecular fragmentation with MP2 perturbation theory, this study presents an innovative approach that enables biomolecular-scale \emph{ab initio} molecular dynamics (AIMD) simulations at WF theory level. Leveraging the resolution-of-the-identity approximation for Hartree-Fock and MP2 gradients, our approach eliminates computationally intensive four-center integrals and their gradients, while achieving near-peak performance on modern GPU architectures. The introduction of asynchronous time steps minimizes time step latency, overlapping computational phases and effectively mitigating load imbalances. Utilizing up to 9{,}400 nodes of Frontier and achieving 59\% (1006.7 PFLOP/s) of its double-precision floating-point peak, our method enables us to break the million-electron and 1 EFLOP/s barriers for AIMD simulations with quantum accuracy. 

\end{abstract}

\begin{IEEEkeywords}
chemistry, exascale, quantum, AIMD, GPU
\end{IEEEkeywords}

\section{Justification for ACM Gordon Bell Prize}
Largest AIMD simulation to date employing quantum-accurate, MP2 wave-function potentials, involving 2{,}043{,}328 electrons; $>$1000$\times$ larger in system-size than state-of-the-art. 
Fastest time-to-solution at this accuracy of 3.4 s/timestep for a 5{,}504-electron protein; $>$1000$\times$ faster than state-of-the-art.
Unprecedented sustained full double-precision performance of 1006.7 PFLOP/s (59\% of Frontier's FP64-peak) for a computational-chemistry application.

\section{Performance Attributes}
\vspace{-0.2cm}
\begin{table}[h]
\centering
\caption{Summary of Performance Attributes}
\begin{tabular}{@{}ll@{}}
\toprule
Performance attribute      & This submission                                      \\ \midrule
Category of Achievement    & Scalability, peak performance, time-to-solution              \\
Type of Method Used        & MBE3/RI-MP2 \emph{ab initio} molecular dynamics                 \\
Results Reported Based On  & Whole application including I/O               \\
Precision Reported         & Double precision                           \\
System Scale               & Results measured on full-scale system      \\
Measurement Mechanism      & Timers, FLOP count           \\ \bottomrule
\end{tabular}
\end{table}

\section{Overview of the Problem
\label{sec:overview}} 

Accurate predictions of molecular characteristics and functionalities are essential for addressing a wide range of societal challenges. These include developing therapeutic drugs, producing biofuels, recycling plastics, and engineering medical biomaterials. Classical molecular dynamics (MD) simulations are pivotal for understanding the chemical and physical properties of biomolecules but are limited by the accuracy of force fields. These parameterized models, which are fitted to experimental and computational quantum chemistry reference data, often inadequately represent complex phenomena like bond breaking and formation, hydrogen bonding, solvent effects, and non-covalent interactions, critical for accurately modeling biomolecular energetics and behavior \cite{D2CS00133K}.

Recent efforts to overcome these limitations have turned to deep learning and machine-learned force fields. Despite progress, challenges such as the scarcity of high-quality training data and the models' limited accuracy and interpretability remain \cite{noauthor_for_2023}.

In contrast, \emph{ab initio} molecular dynamics (AIMD) simulations offer a more accurate and generalizable approach by calculating forces directly from electronic structure theories. However, for AIMD to be practically useful, it must achieve high quantum mechanical accuracy, with absolute energy errors $\le$ 4 kJ/mol in relative energies (\emph{e.g.}, reaction energies),  while jointly being capable of scaling to biologically relevant system sizes with thousands of atoms. The primary challenge in realizing this goal lies in the computational demands of accurately solving the Schr\"{o}dinger equation for large molecular systems. This involves the use of density functional theory (DFT) or wave-function (WF) theory methods, both of which have their advantages and limitations. DFT, for instance, offers a more scalable solution with algorithms developed to scale as $\mathcal{O}(N^3)$ or even $\mathcal{O}(N)$, where $N$ is the number of electrons within the system. 
However, even the most accurate in the class of low-scaling DFT approaches, which utilize hybrid DFT theory, rely on semi-empirical approximations that fall short in yielding the required accuracy \cite{Mardirossian2107}. Notably, this type of DFT struggles with the accurate description of non-covalent interactions, such as van der Waals forces, hydrogen bonding, and dispersion forces \cite{Goerigk2011,Mardirossian2107}, which play a critical role in the chemistry of biomolecular systems \cite{D2CS00133K}.

On the other hand, post-Hartree-Fock (HF) WF-based methods, such as second-order Møller-Plesset perturbation theory (MP2) and its spin-component scaled version SCS-MP2, provide enhanced accuracy, especially for modeling non-covalent interactions \cite{Schwabe2008,Goerigk2011}. 
However, the computational cost associated with MP2, which scales as a prohibitive 
$\mathcal{O}(N^5)$ with respect to the system size, presents a significant challenge.
Efforts to mitigate the computational burden of combined HF plus MP2 calculations have been persistent, and linear scaling MP2 methods have been developed (see Ref.~\cite{Pinski2015} and therein). Yet, the significant computational overhead of these approaches and the inefficiencies within their underpinning algorithms in harnessing large-scale parallel computing resources have traditionally made MP2 calculations unsuitable for bio-scale chemical modeling.

Furthermore, the emphasis on static energy calculations of all the aforementioned approaches, while valuable, falls short of capturing the dynamic nature of biomolecular systems. Dynamic simulations, necessary for predicting macroscopic properties of these systems, require more sophisticated algorithms capable of calculating quantum mechanical energy gradients with efficiency and at scale. 

\begin{figure}[b]
     \vspace{-0.5cm}
     \centerline{\includegraphics[width=0.9\linewidth]{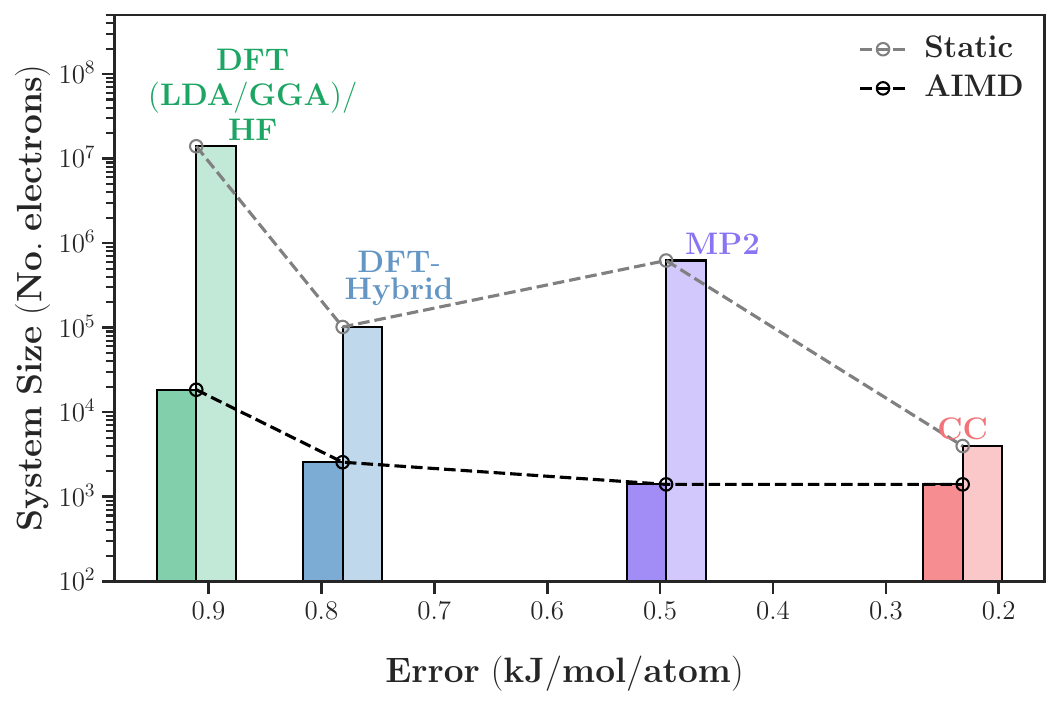}}
    \vspace{-0.4cm}
\caption{Maximum system sizes of static energy evaluations and AIMD simulations achieved with varying levels of theory and corresponding average isomerization energy errors. Isomerization energy errors are obtained from Ref. \cite{Grimme2007}. Exact sizes and corresponding references are listed in Table~\ref{tab:record_table}.}
\label{fig:figure1}
\end{figure}

This trade-off between accuracy and molecular size is illustrated in Fig. \ref{fig:figure1}. The figure compares the maximum number of electrons modeled against the average error in isomerization energies per atom---a reliable indicator of relative energy accuracy---across various DFT and WF methods. This comparison includes both static energy evaluations and AIMD simulations.

\begin{table*}[htbp]
\caption{Largest static energy calculations and AIMD simulations performed at various levels of theory and the corresponding benchmark systems (as shown in Fig.~\ref{fig:figure1}). Basis sets are indicated where applicable.
\vspace{-0.5cm}
}
\begin{center}
\begin{tabular}{>{\centering\arraybackslash}m{0.1\textwidth}>{\centering\arraybackslash}m{0.12\linewidth}>
{\centering\arraybackslash}m{0.08\linewidth}>
{\centering\arraybackslash}m{0.12\linewidth}>{\centering\arraybackslash}m{0.03\linewidth}>{\centering\arraybackslash}m{0.12\linewidth}>
{\centering\arraybackslash}m{0.08\linewidth}>
{\centering\arraybackslash}m{0.12\linewidth}>{\centering\arraybackslash}m{0.03\linewidth}}
\Xhline{1pt}
\multicolumn{1}{c}{\textbf{Level of}} &\multicolumn{4}{c}{\textbf{Static}} & \multicolumn{4}{c}{\textbf{AIMD}} \\
\multicolumn{1}{c}{\textbf{Theory}} & Benchmark System & Basis set & Features & Ref. & Benchmark System & Basis set & Features & Ref.\\
\Xhline{1pt}
\rowcolor[HTML]{c2e9d8}
DFT(LDA/ GGA)/HF & Bulk silicon ($14 \times 10^6$e$^{-}$) & Planewave & Local Orbital approach & \cite{Nakata2020} & Bulk methanol ($18{,}432$e$^{-}$)& MOLOPT- DZVP-SR-GTH & Orbital Transformation optimization & \cite{Taherivardanjani2022}\\

\rowcolor[HTML]{bfd8eb}
DFT (Hybrid) & Bulk water ($101{,}920$e$^{-}$) & - & RI and Numeric Atom-centered Orbitals& \cite{kokott2024efficient} & Bulk water ($2{,}560$e$^{-}$) & Planewave & Maximally localized Wannier functions & \cite{Ko2020}\\

\rowcolor[HTML]{d3c8fb}
 & Ionic liquid cluster ($623{,}016$e$^{-}$) & cc-pVDZ & RI and Fragmentation & \cite{Barca2022} & Bulk water ($1{,}400$e$^{-}$) & aug-cc-pVDZ & Fragmentation & \cite{Liu2017}\\
\cline{2-9}
\rowcolor[HTML]{d3c8fb}
\multirow{-3}{*}{\parbox{0.1\textwidth}{\centering MP2}} & \textbf{Urea cluster (2{,}043{,}328e$^{-}$)} & \textbf{cc-pVDZ} & \textbf{RI and Fragmentation} & \textbf{This work} & \textbf{Urea cluster (2{,}043{,}328e$^{-}$)} & \textbf{cc-pVDZ} & \textbf{RI and Fragmentation} & \textbf{This work} \\

\rowcolor[HTML]{fbc8ca}
CC & Lipid transfer protein ($3{,}980$e$^{-}$) & def2-QZVP & Local Orbital approach & \cite{Nagy2019} & Bulk water ($1{,}400$e$^{-}$) & aug-cc-pVDZ & Fragmentation & \cite{Liu2018}\\
\Xhline{1pt}
\end{tabular}
\label{tab:record_table}
\end{center}
\vspace{-0.5cm}
\end{table*}

Static energy calculations employing hybrid DFT and MP2 methods have been successfully applied to systems containing as many as 101{,}920 and 623{,}016 electrons, respectively \cite{kokott2024efficient, barca2021enabling, Barca2022}. In contrast, Fig.~\ref{fig:figure1} shows that owing to their higher algorithmic complexity, AIMD simulations have been constrained to comparatively smaller systems. The largest hybrid DFT and MP2 AIMD simulations have been executed on systems with 2{,}560 and 1{,}400 electrons, respectively \cite{Ko2020, Liu2017} (see Table~\ref{tab:record_table}).

Thus, there is an absence of AIMD methodologies that utilize high-fidelity, WF-based \emph{ab initio} potentials, such as MP2, to enable high-accuracy biomolecular simulations.

\subsection{Summary of Contributions}

This work presents innovative algorithmic advancements that, for the first time, allow for the execution of large-scale AIMD simulations utilizing MP2 potentials, enabling the study of biomolecular-scale systems with unprecedented accuracy. 

The main algorithmic innovations from this work are:
\begin{enumerate}
\item[(i)] An efficient multi-layer scheme for distributed many-GPU AIMD calculations using molecular fragmentation at the third-order many body expansion (MBE3) level. This reduces the scaling of MP2 calculations from quintic to linear with system size while retaining quantum accuracy. 
\item[(ii)] A  fully electronically correlated gradient implementation, based on the resolution-of-the-identity (RI) approximation \cite{Weigend1997}, that combines RI-HF and RI-MP2 in a synergistic manner. This approach eliminates expensive four-center electron repulsion integrals, avoids the recomputation of expensive three-centre integrals, and replaces the traditional, computationally-inefficient bottlenecks of these methods with efficient sequences of dense matrix multiplications.
\item[(iii)] An asynchronous time step AIMD scheme that leverages the dependencies among fragments to asynchronously allow a subset of the system to progress to the next time step while the remainder of the previous time step is completed. This significantly enhances distributed workload balancing with small time step latencies of the order of individual fragment latency.
\item[(iv)] A matrix multiplication auto-tuning scheme for performance portability across different machines that determines and implements the highest-performance DGEMM configurations at runtime.
\end{enumerate}

This novel approach enables us to perform several AIMD time steps on a molecular cluster with over 2 million electrons utilizing 9{,}400 nodes on Frontier, significantly larger than any previous AIMD or static energy and/or gradient calculation at a comparable level of accuracy. These calculations achieve 1006.7 PFLOP/s providing a throughput efficiency of 59\% of attainable FP64 peak on 99.9\% of the machine. In addition, we demonstrate low time step latency of 3.4 s/timestep on a protein fragment with 1{,}496 atoms and over 5.5k electrons attaining a simulation throughput of 25{,}000 time steps per day on 1{,}024 Perlmutter nodes.

\section{Current State of the Art} \label{sec:sota}

In this section, we provide an overview of significant developments in scalable quantum mechanical approaches for static energy calculation and AIMD simulations. A subset of the studies discussed herein are summarized in Table~\ref{tab:record_table}.

The DFT framework has facilitated the development of inherently more scalable approaches than those derived from WF theory, including algorithms with $\mathcal{O}(N^3)$ and $\mathcal{O}(N)$ complexities. However, the accuracy of DFT approaches relies on the usage of an exact exchange-correlation (XC) functional to encapsulate electron-electron interactions. Despite the theoretical existence of a universal XC functional, DFT lacks a systematic approach for its determination, rendering it in practice a semi-empirical theory. This gap has led to the empirical development of a multitude of approximate XC functionals, guided by physical insights and statistical performance evaluations.

DFT approaches are classified in five rungs of an increasing accuracy ladder: 1) local density approximation (LDA) DFT, 2) generalized gradient approximation (GGA) DFT, 3) meta-GGA, 4) hybrid GGA/meta-GGA DFT  and 5) double-hybrid DFT\cite{mardirossian2017thirty}.
Unfortunately, no scalable methods have been developed for the fifth rung double hybrids which require a constituent MP2 calculation. Specifically, to date, the most accurate large-scale DFT calculations use hybrid DFT. The FHI-aims code recently performed calculations on a 10{,}192 molecule ice system (101,920 electrons) but took approximately 40 minutes per SCF iteration~\cite{kokott2024efficient}.
On the other hand, Ko \emph{et al.} performed AIMD simulations on condensed-phase water systems comprising up to 256 molecules (2{,}560 electrons) at the hybrid PBE0 level of theory \cite{Ko2020}. As shown in Fig. \ref{fig:figure1}, due to its parameterized nature, all levels of DFT approximation up to hybrid DFT remain far from the accuracy of post-HF WF theory.

In contrast, post-HF WF methods, such as perturbation theory, including MP2, and coupled-cluster (CC) theories, provide superior accuracy in calculating quantum molecular energies and gradients.  However, their practical application to large molecules is hindered by their steep computational scaling of $\mathcal{O}(N^5)$ for MP2 and $\ge\mathcal{O}(N^6)$ for CC calculations.

Significant research has focused on developing more efficient algorithms and software for MP2 and CC energy evaluation. Among these, the resolution-of-the-identity (RI) approximation and the RI-MP2 method \cite{Weigend1997} have become prominent for their ability to accelerate calculations with minimal error introduction. Further efforts to mitigate the steep computational scaling of MP2 and CC energies and gradients have led to the development of various lower-order scaling algorithms \cite{Pinski2015}. These exploit the local nature of electronic correlation through strategies like orbital localization, atomic-level truncation, matrix element sparsity, or molecular fragmentation, enabling their application to larger molecules.

Notable achievements of scalable methods for MP2 include: 1) In 2016, a fragment molecular orbital level 3 (FMO3) MP2 gradient calculation using the 6-31G(d,p) basis was performed on 64 water molecules using 4{,}096 nodes (65,536 CPU cores) of the Mira Blue Gene/Q supercomputer with 45\% parallel efficiency; no dimer or trimer distance cutoffs were reported\cite{pruitt_importance_2016}. 2) In 2016, a divide-expand-consolidate RI-MP2/cc-pVDZ/cc-pVDZ-RIFIT calculation was used to calculate the energy of 1-aza-adamantane-trione supramolecular wires containing up to 2{,}440 atoms, as well as an RI-MP2 gradient simulation of insulin (787 atoms) on the Titan supercomputer \cite{bykov_molecular_2016,thomas_k_2017, bykov_gpu-enabled_2017}. 3) In 2019, FMO2/RI-MP2 single point gradient calculations with the 6-31G(d,p) basis set on 6,495 atoms using up to 768 nodes of the Theta supercomputer \cite{pham_hybrid_2019,pham_development_2020}, which appears to be the high watermark for MP2 static gradient calculations; no full HF plus MP2 nor MP2 timings were reported. 4) In 2021, various FMO2 correlated static energy calculations on the SARS-CoV-2 spike protein with 23,694 atoms using  3,072 nodes and 147,456 cores of the Fugaku supercomputer, with wall times of the order of hours \cite{akisawa2021interaction}; 5) In 2022 an FMO2/RI-MP2 static energy calculation at the 6-31G(d)/cc-pVDZ-RIFIT level on an ionic liquid crystal lattice structure with 146{,}592 atoms was performed by us on Summit \cite{Barca2022}.

\begin{table}[htbp]
\centering
\vspace{-0.2cm}
\caption{Single time step latency (s) for AIMD/RI-MP2/cc-pVDZ (including HF component) of varying length polyglycines (Gly{$_n$}) using common quantum chemistry software packages.}
\label{tab:gradient_calculations}
\setlength{\tabcolsep}{3pt} \setlength{\extrarowheight}{-3pt}
\begin{tabular}{
    c 
    | S[table-format=4.0,table-align-text-post=false] 
    S[table-format=4.0,table-align-text-post=false] 
    S[table-format=4.0,table-align-text-post=false] 
    S[table-format=4.0,table-align-text-post=false] 
    | S[table-format=3.2,table-align-text-post=false] 
    S[table-format=3.2,table-align-text-post=false] 
    S[table-format=3.2,table-align-text-post=false] 
}
\toprule
\multirow{2}{*}{} & {\makecell{Orca}} & {\makecell{Q-Chem}} & {\makecell{GAMESS}} & {\makecell{NWChem}} & \multicolumn{3}{c}{\textbf{This work (EXESS)}} \\
\cmidrule(lr){2-8}
& \multicolumn{5}{c|}{No fragmentation} & \multicolumn{2}{c}{MBE3} \\
\cmidrule(lr){2-6}
\cmidrule(lr){7-8}
Gly$_{n}$ & \multicolumn{4}{c|}{\makecell{nCPU=2, ncore=104 \\Sapphire Rapids}}   & \multicolumn{1}{c|}{\makecell{4$\times$\\ A100}} & {\makecell{4$\times$ \\ A100}} & {\makecell{16$\times$ \\ A100}} \\
\midrule
10 & 297  & 252   & 258 & 1477  & 6 &  2.7 & 1.1 \\
15 & 1976 & 1050 & 1573 & {--}  & 24 & 4.4 & 1.4 \\
20 & 6213 & 5710  & {--} & {--}  & 83 &  6.4 & 1.6  \\
\bottomrule
\end{tabular}
\vspace{-0.5cm}
\end{table}

Although static calculations at the MP2 level have been steadily increasing in size, dynamic calculations have lagged. As listed in Table~\ref{tab:record_table}, \textbf{the high watermark for molecular sizes achieved in AIMD simulations using fragmentation with MP2 potentials was performed on a 140 water cluster system (1{,}400 electrons) using the aug-cc-pVDZ basis set} \cite{Liu2017}. The authors comment on the latency per time step as being large but provide no measure of it \cite{Liu2017}. To establish an effective timing benchmark for AIMD simulations using MP2 potentials, Table \ref{tab:gradient_calculations} shows a comparative analysis of the wall time of Hartree-Fock plus RI-MP2 gradient calculations, which provides the single AIMD time step latency, across state-of-the-art quantum chemistry software packages~\cite{neese_orca_2012,GAMESS,qchem4,apra_nwchem_2020}. For the MBE3 calculations, the RI-HF approximation was employed for the Hartree-Fock component, with glycine chains fragmented into monomers composed of individual amino acids. The dimer and trimer cutoffs for these MBE3/RI-MP2 calculations were set at 20 \AA{} and 13 \AA{}, respectively, which were obtained by considering all dimers and trimers with absolute energy contributions to the total molecular energy greater than $0.1$ kJ/mol. Compared to the non-fragmented calculation, this yields a gradient RMSD of less than $10^{-4}$ Hartree/Bohr, which is commonly adopted as a geometry optimization convergence threshold.
Note that to the best of our knowledge EXESS is the only software reported in the literature supporting analytic RI-MP2 gradients on GPUs. \textbf{As shown in Table \ref{tab:gradient_calculations}, in combination with the MBE3 fragmentation framework described in later sections, this work is faster than state-of-the-art CPU software by three orders of magnitude already on a single node (4$\times$A100). Below we demonstrate significant additional speedups with increasing node counts}.

Moreover, both static and dynamic MP2 modelings at scale typically fall short in accuracy due to either the absence of a sufficiently large basis sets (a minimum requirement being a double-$\zeta$ polarized basis set such as cc-pVDZ) or because all calculations, when employing fragmentation, were conducted at the dimer level, accompanied by cutoffs that prematurely and inaccurately truncate long-range interactions. For example, in Ref. \cite{Liu2017} an unjustified short dimer distance cutoff of 5 \AA{} is used, and trimers are not included (more on this in Section \ref{sec:molfrag}). 

In summary, the current state of the art in AIMD methods is limited by either the accuracy of the method used to calculate the forces, or the cost with respect to system size for the \textit{ab initio} calculation. The framework described herein aims to bridge the gap between the scalability and accuracy dilemma through the use of molecular fragmentation and tailor made quantum chemistry programs. Our study shows that when the RI-HF and RI-MP2 methods are implemented in an optimal fashion, coupled with a rigorous and accurate molecular fragmentation scheme, AIMD at the correlated level of theory is possible at the scale of  $\mathcal{O}(10^{4-6})$ electrons.

\section{Innovations Realized}\label{sec:innov_realized}

\subsection{Notation}\label{sec:notation}

We will adopt the following notation throughout. Indices $i, j, k$  denote occupied molecular orbitals, whilst $a, b, c$ denote virtual orbitals and $p, q, r, s$ index over all molecular orbitals (occupied and virtual). Primary and auxiliary basis functions are denoted $\mu,\nu,\lambda,\sigma$ and $P, Q, R$ respectively. Fragment monomers are indexed $I, J, K$. Sums involving any of these indices are implicitly over the full range of the index. Chemist's notation is used for electron repulsion integrals (ERIs) where the vertical bar $|$ denotes the Coulomb repulsion operator $\frac{1}{r_{12}}$, for example the three-center integral
\begin{equation}
    \begin{aligned}
        (\mu\nu|P)=\int{\psi_\mu}(\boldsymbol{r}_{1})&\psi_\nu(\boldsymbol{r}_1) \frac{1}{|\boldsymbol{r}_1-\boldsymbol{r}_2|}\psi_P(\boldsymbol{r}_2)d\boldsymbol{r}_1d\boldsymbol{r}_2.
    \end{aligned}
\end{equation}
where $\{\psi_\mu\}$ are orbital functions in the primary/auxiliary atomic orbital (AO) or molecular orbital (MO) basis. 

In all sections hereafter, we assume the restricted, closed shell formulation of HF and MP2, and do not use the frozen core approximation.

\subsection{Molecular Fragmentation
\label{sec:molfrag}
}

We adopt a molecular fragmentation approach whereby the overall system is divided into individual fragments, called monomers. The energy and gradient of these fragments can be individually calculated and gathered to approximate the original, unfragmented system, according to the following many-body expansion (MBE)
\begin{equation}
    \label{eq:mbe3}
    E = \sum_I E_I + \sum_{I < J} \Delta E_{IJ} + \sum_{I<J<K} \Delta E_{IJK} + ...
\end{equation}

The MBE method breaks down the system's energy into individual monomer energies and corrections from pairwise and higher-order interactions. In this framework, $E_I$ represents a monomer's energy, and $\Delta E_x$ accounts for higher-order corrections. To prevent double counting, we subtract lower-order contributions from higher ones. For instance, the pairwise correction $\Delta E_{IJ} = E_{IJ} - E_I - E_J$ involves removing monomer energies $E_I$ and $E_J$ from the dimer energy $E_{IJ}$. When the system requires breaking a single bond, we add a hydrogen cap (H-cap) to maintain chemical integrity, replacing the bonded atom in isolated monomers.

Hereafter, we use the term \emph{polymer} to denote a group of monomers forming a cohesive unit. Research demonstrates that the truncated Eq.~\eqref{eq:mbe3} at second and third orders, considering polymers within a threshold distance \(R_{\text{cut}}\), yields accurate energy approximations of the original system using MBE2 and MBE3 with linear computational costs \cite{gordon_fragmentation_2012}. However, selecting an appropriate \(R_{\text{cut}}\) for large systems can be arbitrary and imprecise. In response, this study conducts MBE calculations centered around a reference monomer to accurately determine the energy contributions from associated polymers, ensuring a precise method for setting \(R_{\text{cut}}\) with minimal accuracy loss. Additionally, achieving quantum mechanical precision within 2~kJ/mol/monomer for scaled molecular systems necessitates the use of MBE3 \cite{herbert_fantasy_2019}, supported by our recent findings on the precision of MBE3 RI-MP2 gradients \cite{Stocks2024}. Consequently, our approach utilizes an MBE3 expansion to calculate the energy and gradient of each polymer, incorporating the Hartree-Fock and MP2 corrections with the RI approximation as follows

\begin{align}
    E_{f} & = E_{f}^{RI-HF} + E_{f}^{RI-MP2} \\
     \nabla{E}_{f} & = \nabla{E}_{f}^{RI-HF} + \nabla{E}_{f}^{RI-MP2}
       \label{eq:efrag}
\end{align}
where $f = I, IJ, IJK$. 

\begin{figure*}[t]
    \centering
    \includegraphics[width=1\textwidth]{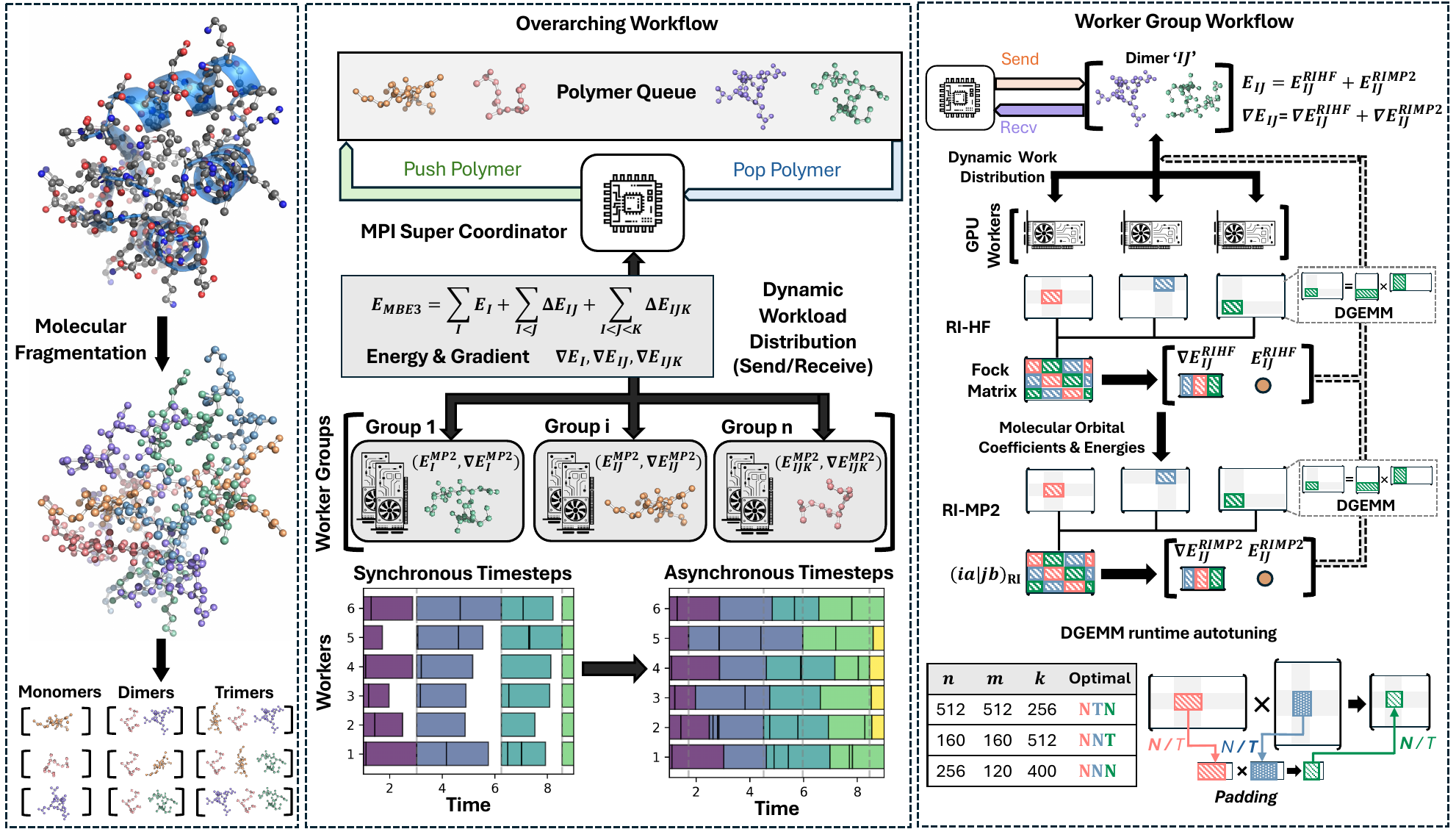}
    \caption{
     \vspace{-0.8cm}
    Overarching algorithmic scheme
 adopted for the AIMD/RI-HF+RI-MP2 calculations.
 \vspace{-3cm}}
     \label{fig:overaching_algo}
\end{figure*}

\subsection{
\label{subsec:RIHFMP2}
Resolution-of-the-Identity HF and MP2
}

We adopt the RI approximation~\cite{Vahtras1993} for both HF and MP2. This is based on the observation that products of primary basis functions can be expressed as a linear combination of auxiliary basis functions. Under this approach, the four-center two-electron repulsion integrals can be approximated, with negligible loss of accuracy, by a combination of two and three-center integrals
\begin{equation}\label{eq:symmetric_ri}
	{(\mu\nu|\lambda\sigma)}\approx{(\mu\nu|\lambda\sigma)}_{RI} = \sum_{P} B_{\mu\nu}^P B_{\lambda\sigma}^P,
\end{equation}
where
\begin{equation}\label{eq:biap}
    B_{\mu\nu}^P = \sum_{Q} (\mu\nu|P) J^{-\frac{1}{2}}_{PQ}.
\end{equation}

$J_{PQ}^{-\frac{1}{2}}$ denotes the matrix inverse square root of the positive definite two-center integrals $(P|Q)$. This can also be directly transformed to the molecular orbital (MO) basis $B_{ia}^P=\sum_{\mu\nu} C_{i\mu} C_{a\nu} B_{\mu\nu}^P$.

This formulation enables the computational bottlenecks of HF and MP2 calculations to be recast as sequences of matrix multiplications that can be executed at near-peak floating point performance, thereby significantly reducing the computational pre-factor of these bottlenecks through a more efficient utilization of the hardware. 

For example, the bottleneck of the traditional formulation of HF is the calculation of $\mathcal{O}(N^4)$ four-center electron repulsion integrals $(\mu\nu|\lambda\sigma)$\cite{galvez2023high} and their ensuing combination with the density matrix $D_{\lambda\sigma}$ to form Fock matrix elements
\begin{equation}\label{eq:fock_build}
    F_{\mu\nu} =  \sum_{\lambda \sigma} D_{\lambda\sigma} [(\mu\nu|\lambda\sigma)- \frac{1}{2} (\mu\sigma|\lambda\nu)].
\end{equation}

When performed in parallel this is typically an inefficient process due to the following reasons:
\begin{itemize}
 \item The calculation of $(\mu\nu|\lambda\sigma)$ involves different workloads, which can be memory bound and run at low FLOP rates depending on the specific nature of the AO basis functions involved. This creates workload imbalance and FLOP-efficiency issues for parallel execution. 
 \item There are too many $(\mu\nu|\lambda\sigma)$  integrals to be stored in memory, so they are often recomputed at each iteration of the self-consistent field (SCF)  Hartree-Fock algorithm. 
 \item To save on the calculation of the integrals, permutational symmetry is used. In turn, this causes scattered memory access patterns (both reads and writes) and potential race conditions when the $(\mu\nu|\lambda\sigma)$  are combined with $D_{\lambda\sigma}$ and the result stored in the Fock matrix $\mathbf{F}$. This renders this stage usually memory bound and with low bandwidth utilization on throughput-oriented architectures such as GPUs, thereby further lowering algorithmic performance. 
\end{itemize}

In contrast, using the RI approximation in Eq. \eqref{eq:symmetric_ri}, one can:
\begin{itemize}
\item Compute only $\mathcal{O}(N^3)$ three-center integrals $(\mu\nu|P)$ and $\mathcal{O}(N^2)$ two-center integrals $(P|Q)$.
\item These integrals can be computed once and stored in host or even GPU memory.
\item The formation of the Fock matrix can then be performed according to the following equation 
\begin{equation}\label{eq:fock_build_RI}
    F_{\mu\nu} = \sum_{P} \sum_{\lambda \sigma} D_{\lambda\sigma} [B_{\lambda\sigma}^{P}B_{\mu\nu}^{P}- \frac{1}{2} B_{\lambda\nu}^{P}B_{\mu\sigma}^{P}]
\end{equation}
The bottleneck of this formulation can be implemented using DGEMMs to perform the contraction of slices of the $\mathbf{B}$ tensor. This is now a memory and FLOP-efficient process, especially for GPU architectures. 
\end{itemize}

Similar considerations apply to the MP2 algorithm, whose bottleneck under the RI formalism becomes the formation of the two-electron integrals in the MO basis
\begin{equation}\label{eq:MP2_ints}
    (ia|jb)_{RI} = \sum_{P} B_{ia}^P B_{jb}^P
\end{equation}
which can also be implemented using DGEMMs.

\subsection{Overarching Algorithm}

An overview of our overarching algorithm is shown in Fig. 2.
Our implementation uses a multi-layer dynamic load distribution scheme utilizing MPI in which a centralized super coordinator distributes polymers to worker groups from a centralized polymer priority queue. Workers then compute the RI-MP2 energy and gradient of their assigned polymers on GPUs, primarily utilizing dense linear algebra operations as a result of the RI formulation. The full RI-MP2 gradient calculation including the calculation of the two and three-center integrals is performed on the GPUs to avoid CPU bottlenecks ~\cite{GalvezVallejo2023,Palethorpe2024}. The resulting energy and gradient is then returned to the coordinator alongside the request for a new work allocation. The coordinator then stores the energy and gradient and determines which monomers are ready to be updated to the next time step. These updates are performed using velocity Verlet time step integration.

Worker groups are confined to a single node and can utilize any number of GPUs within the node. There can also be several worker groups per node, each utilizing one or more GPUs. Each worker group has one rank per GPU with an additional local coordinator rank to enable dynamic load distribution within the group.

The following sections will describe in more detail the synergistic RI-HF and RI-MP2 gradient formulation, the asynchronous time step behavior, and our runtime auto-tuning to optimize DGEMM routines.

\subsection{Synergistic RI-HF and RI-MP2 gradient formulation}

\begin{figure}[t]
    \vspace{-0.4cm}\centerline{\includegraphics[width=0.9\linewidth]{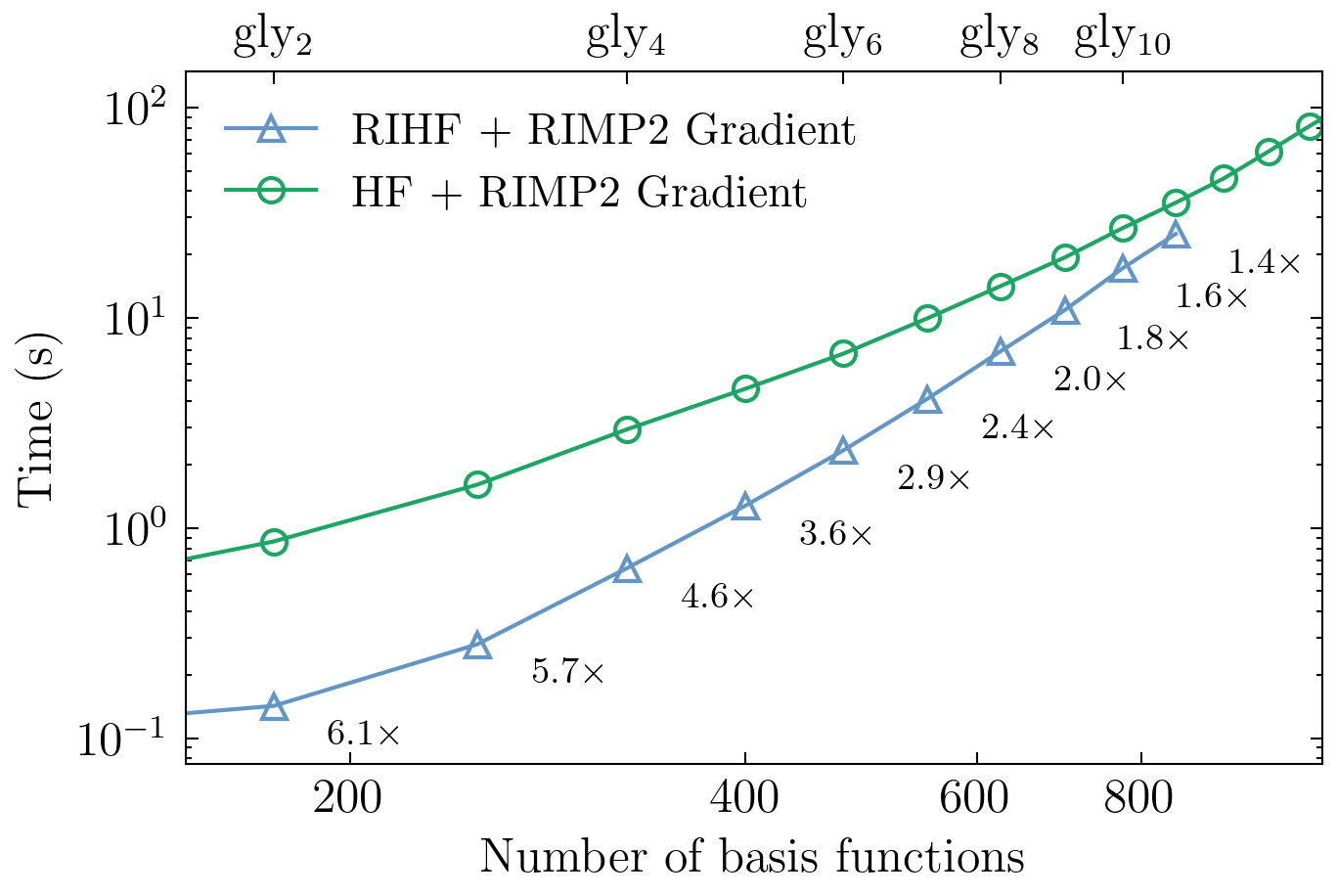}}
    \vspace{-0.1cm}
\caption{
Execution time of RI-MP2 gradients with and without the RI-HF approximation for varying length glycine chains on a single 40 GB NVIDIA A100 GPU\@. The cc-pVDZ/cc-pVDZ-RIFIT primary and auxiliary basis set were used. Data points are labelled with the speedup of the RI vs non-RI HF version. }
\vspace{-0.5cm}
\label{fig:ri_grads}
\end{figure}

In molecular dynamics simulations, reducing the cost of each time step is crucial. The fragmentation framework helps by distributing the computational load across multiple nodes. However, to decrease latency, it is also necessary to minimize the computational cost of each fragment. Traditionally, computational chemistry algorithms are optimized for large molecular fragments that require hours or days per computation. For quicker time step processing, smaller fragments are preferable to reduce latency to mere seconds. For example, although Stocks \emph{et al.} achieved over 80\% of peak floating point throughput calculating RI-MP2 gradients, these calculations took more than 30 minutes per fragment \cite{Stocks2024, GalvezVallejo2023}, making them impractical for dynamic simulations. Additionally, for smaller systems, the computation of four-center electron repulsion integrals, especially the gradients, becomes a significant bottleneck.

\newcommand{\epsij}[1][]{\epsilon_{i #1} + \epsilon_{j #1}}
\newcommand{\epsai}[1][]{\epsilon_{a #1} - \epsilon_{i #1}}
\newcommand{\epsab}[1][]{\epsilon_{a #1} - \epsilon_{b #1}}
\newcommand{\epspq}[1][]{\epsilon_{p #1} + \epsilon_{q #1}}

In this work, we extend the RI-MP2 gradient formulation presented by Weigend and H{\"a}ser~\cite{Weigend1997} to additionally use the RI approximation for the HF component, eliminating the need for four-center integrals and their gradients entirely~\cite{Stocks2024a}. 

The analytic RI-MP2 corrected gradient with respect to a geometric perturbation $\xi$ becomes
\begin{align}\label{eq:rimp2_gradient}
    E^\xi_{HF+MP2}=&\sum_{\mu\nu P} Z_{\mu\nu}^P{(P|\mu\nu)}^\xi + \sum_{PQ} \zeta_{PQ}{(P|Q)}^\xi\\
                   &+ 2\sum_{\mu\nu}D_{\mu\nu}h_{\mu\nu}^\xi-2\sum_{\mu\nu}W_{\mu\nu}S_{\mu\nu}^\xi,
\end{align}
where $S_{\mu\nu}^{\xi}$ and $h_{\mu\nu}^{\xi}$ represent the first derivative of the overlap matrix and core Hamiltonian, respectively. The remaining intermediates are calculated through a series of linear algebra operations (primarily DGEMMs) as defined in the Appendix.

This formulation has been specifically designed such that the full coefficients of the integral derivatives ($ Z_{\mu\nu}^P, \zeta_{PQ}, D_{\mu\nu}, $ and $W_{\mu\nu}$) are computed first. This allows the electron repulsion integral derivatives to be calculated and accumulated into the final gradient on the fly, without needing to be stored. Due to targeting small fragments, the three-center integrals are stored directly in GPU memory. While this limits achievable individual fragment sizes to approximately 1k basis functions per fragment for a 40 GB GPU, it eliminates large CPU-GPU transfers. When multiple GPUs are used for a single fragment, a subset of the three-center integrals are stored on each GPU (batched by the auxiliary index) such that achievable fragment size can be increased by using more GPUs.

Whilst RI-HF formally scales as $\mathcal{O}(N^4)$ relative to $\mathcal{O}(N^3)$ for a direct HF calculation with integral screening, RI-HF has a much smaller pre-factor and can be expressed primarily as a sequence of matrix multiplications rather than complex four-center integrals. For very small fragments, this can provide up to 6$\times$ speedup and is faster than previous implementations where the RI approximation is used only for the MP2 component across the full range of accessible fragment sizes, as shown in Fig.~\ref{fig:ri_grads}.

\subsection{Asynchronous time steps}

A significant challenge when scaling MBE fragmented calculations to a large number of parallel resources is balancing the workload due to the different computational costs of differing sized fragments. Previous single point energy calculations have relied on producing an extremely large number of fragments that can be dynamically distributed to workers such that the total computation time is much larger than that of a single fragment. However, in the context of AIMD, the full MBE gradient must be recomputed each time step and thus alternative methods of load balancing are required to increase efficiency whilst retaining minimal time step latency. 

In this work, we propose a novel method to asynchronously allow a subset of the system to progress to the next time step whilst the remainder of the previous time step is completed. This is achieved with no compromise to accuracy and enables full utilization of computational resources by eliminating all system wide synchronization.

An arbitrary fragment towards an extremity of the system is first chosen to be the reference fragment. Polymers for each time step are then stored in a priority queue ordered first by the minimum distance of one of the constituent monomers to the reference fragment, then ordered 
by decreasing size. This ensures that polymer calculations are ordered such that all polymers involving the monomers near the reference are completed first. This allows these monomers to be integrated to the next time step and begin computation asynchronously while the remainder of the previous time step polymer calculations are completed. Tie-breaking on polymer size ensures that larger polymers with longer compute latency are started first, leaving smaller polymers to fill in any gaps at the end of the time step. 

\begin{figure}[t]
    \centering
    \includegraphics[width=0.45\textwidth]{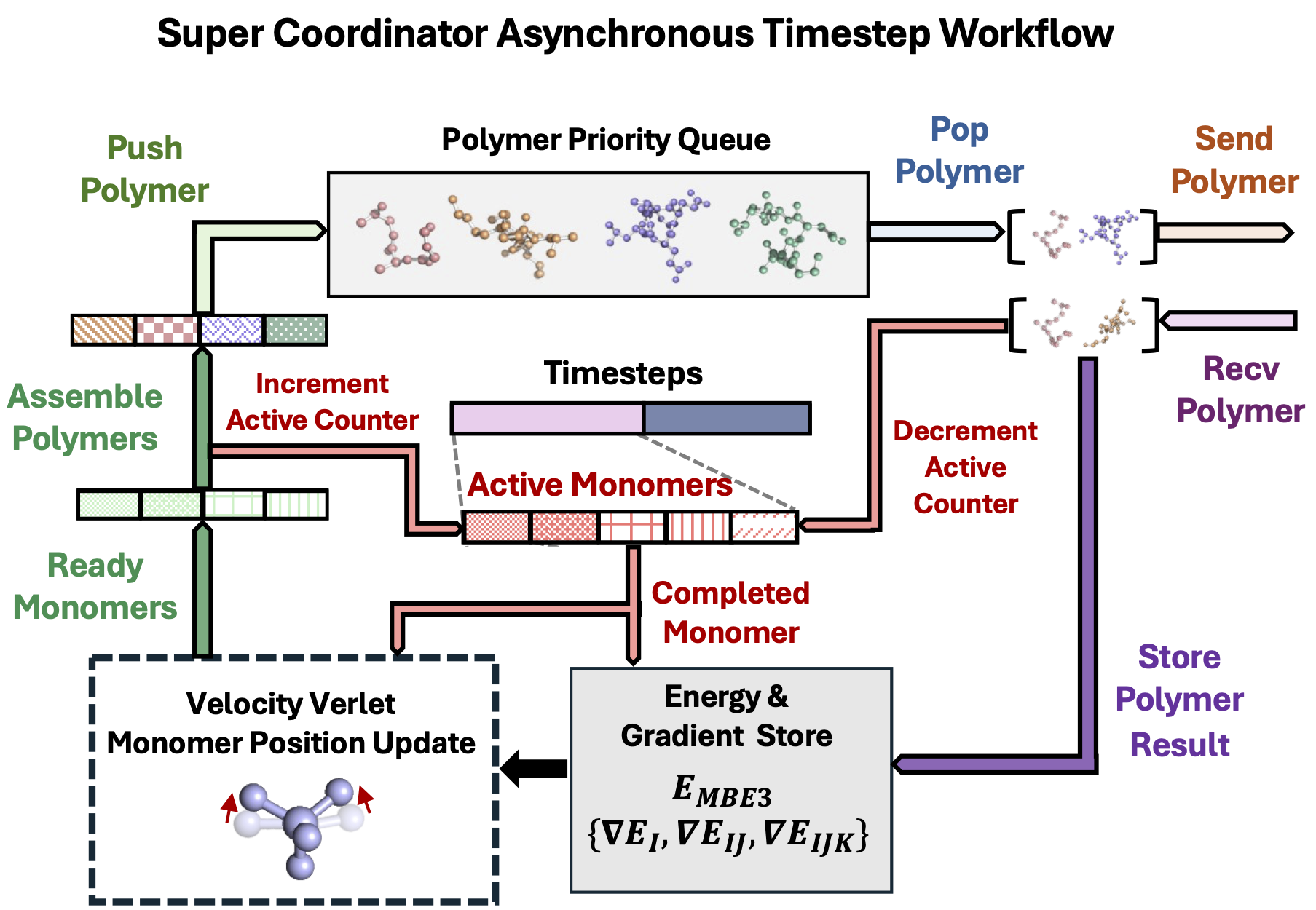}
    \caption{Schematic of the asynchronous time step workflow showing the major algorithmic components and data flow.}
    \vspace{-0.5cm}\label{fig:async_workflow}
\end{figure}

Only polymers that are ready to begin computation are stored in the priority queue so that a worker's request for additional work can be immediately fulfilled by popping a polymer from the queue as shown in Fig.~\ref{fig:async_workflow}. When the energy and gradient are received from a worker, they are accumulated and stored until there are no longer any active polymers that include a particular monomer. That monomer's component of the overall gradient is then calculated using the MBE of the polymer gradients and used to update the monomer atom positions to the next time step. All of the polymers consisting of this monomer and other updated monomers can then be formed and inserted into the queue for the next time step.

Storing all of the trimer gradients independently requires an enormous amount of memory and becomes a bottleneck to the formation of the MBE monomer gradients. However it is guaranteed that the coefficient of all trimer gradients is 1 in the final MBE so these gradients can be accumulated directly into a full system gradient, combining only the monomer and dimer contributions on the fly when a monomer is completed.

Two approaches identify fragments for release based on system size. For small systems, fragments are dynamically constructed each time a monomer updates. This involves searching through all updated monomers to form fragments with the current monomer. For large systems, frequent searches create bottlenecks. Therefore, the full polymer list is generated every few time steps. When a monomer updates, it is added to a pre-formed list, improving efficiency without sacrificing accuracy, assuming the fragment cutoff is adequately adjusted for monomer velocities.

Fragments with broken bonds can also be updated asynchronously, however they cannot be inserted into the queue for the next time step until the surrounding fragments have also been updated so that the H-caps can be placed accordingly. This is maintained as a separate list of dependencies based on the initial bond structure.


\subsection{GEMM runtime auto-tuning}
Vendor-provided general matrix to matrix multiplication (GEMM) routines are known to perform extremely well on GPU accelerated architectures. The GEMM API allows the computation of $AB$, $A^TB$, $AB^T$ or $A^TB^T$. Each of these options may be performed by a different underlying algorithm and, consequently, may exhibit different performance characteristics. This is demonstrated in Table~\ref{tab:dgemm_performance}, which shows a performance difference of up to a factor of 20 between the different algorithmic variants. It should also be noted that the best variant varies on a case-by-case basis and can also vary by machine and library versions.

\begin{table}[t]
\begin{center}
\vspace{-0.2cm}
\caption{Single-GCD DGEMM performance on AMD Instinct\texttrademark{} MI250X GPU (ROCm 5.4.3) on 3 matrix shapes arising in RI-MP2 gradient calculations.
\vspace{-0.1cm}
}
\renewcommand{\arraystretch}{1.05}
\begin{tabular}{ccc||cccc}
\hline
\multicolumn{3}{c||}{\textbf{Matrix Shape}} &\multicolumn{4}{c}{\textbf{Performance (TFLOP/s)}} \\
$m$ & $k$ & $n$ & NN & NT & TN & TT \\
\hline
960 & 324480 & 960 & 15.9 & 19.5 & 14.3 & 11  \\
120 & 2957880 & 120 & 2.92 & 0.332 & 7.12 & 0.370 \\
192 & 738048 & 192 & 6.8 & 0.89 & 9.4 & 0.54 \\
\hline
\end{tabular}
\label{tab:dgemm_performance}
\end{center}
\vspace{-0.5cm}
\end{table}

Given that matrix transpose operations are inexpensive in comparison to GEMM operations, it is possible to change the GEMM variant by transposing one or both of the input matrices. For example, the calculation $C = AB$ can be recast as $D = A^T$ followed by $C = D^TB$. This allows the best-performing variant to be utilized, provided that it is known. 

The runtime auto-tuning system employed in this work dynamically determines the optimal GEMM variant for each GEMM shape encountered during execution, providing performance portability across a range of systems without relying on machine-specific optimizations. For each GEMM shape, the auto-tuning system trials each algorithmic variant, measuring the performance (including any transpose operations) with CUDA/HIP event timers. Once each variant has been tested, the determined best variant is used for all subsequent GEMM calls with the same shape.

The measurement of each variant is performed \textit{in-situ} to avoid redundant work. If a given GEMM shape is performed 10 times in a calculation, the first 4 computations are used to trial each variant while the remaining 6 use the best variant.

Performance improvement from this auto-tuning process can only be expected if the same GEMM shapes are repeated throughout the computation. AIMD is an ideal application for this method as the most expensive GEMM shapes typically appear 10$\times$ to 100$\times$ in each gradient calculation, and this is further multiplied by the number of time steps. The auto-tuning procedure achieved a speedup of 13\% for an AIMD simulation of a urea trimer, and 12\% for a paracetamol trimer on a single AMD Instinct\texttrademark{} MI250X Graphics Compute Die (GCD).

\section{How Performance Was Measured}\label{sec:systems_perf}
\subsection{HPC platforms and software environment}

All calculations were performed on the OLCF Frontier and NERSC Perlmutter supercomputers, ranked number 1 and 14 on the June 2024 TOP500 list, respectively~\cite{Robert2011}.

Frontier is a Cray EX supercomputer with 9{,}408 nodes, each containing an Optimized 3rd Generation AMD EPYC\texttrademark{} 64-core CPU and 4 AMD Instinct\texttrademark{} MI250X GPUs. Each GPU has two GCDs, each with 64 GB of HBM2e memory and a sustainable peak double precision matrix throughput of 22.8 TFLOP/s for a total sustainable peak of the machine of 1.715 EFLOP/s.

Perlmutter is a Cray Shasta machine with 1{,}536 GPU nodes, each containing one AMD EPYC\texttrademark{} 7763 CPU and 4 NVIDIA A100 GPUs. Each GPU has 40 GB of HBM2 memory and a theoretical peak double precision matrix throughput of 19.5 TFLOP/s (18.4 TFLOP/s sustained) for a total sustainable peak of the machine of 113 PFLOP/s.

Both systems are connected by a Slingshot-11 dragonfly interconnect with at most three hops between any two nodes.

\subsection{Benchmark molecular systems}

To assess our program's scalability and computational efficiency, we tested it on increasing-radii spherical sections of crystal lattices from selected biomolecules. These include urea $(\text{C}\text{H}_4\text{N}_2\text{O})$ and paracetamol $(\text{C}_8\text{H}_9\text{N}\text{O}_2)$. These molecules were selected for their significant academic and industrial relevance. For instance, urea's different crystalline forms are utilized across various sectors, including solvent production, and the pharmaceutical and cosmetics industries. Similarly, the crystal lattice stability of paracetamol is crucial for therapeutic applications. These compounds exhibit multiple polymorphs and a predominance of non-covalent interactions, areas where hybrid DFT underperforms \cite{Civalleri2007}. Polymorphism---the ability of materials to exist in multiple crystalline forms---influences bulk properties like solubility, dissolution and drug efficacy. Predicting polymorphs is very challenging, as lattice energy differences between polymorphs are usually under 2 kJ/mol, beyond the accuracy of force fields and hybrid DFT \cite{Nyman2016}.

Recent advances show that combining MBE3 with scaled MP2, can accurately predict lattice energies and thus polymorph energetics \cite{GalvezVallejo2023}. Our program leverages this method to perform calculations that successfully predict the stability of these crystal structures for the first time at the MP2 level.

Moreover, an isolate of the prion-protein (PrP) protein fibril (PDB ID: 6PQ5), was utilized to evaluate the program's energy conservation and time step latency. PrP isolates have been used to establish the molecular mechanisms of most prion diseases, such as fatal familial insomnia\cite{glynn_cryo-em_2020}. The system has 360 atoms across 36 monomers with 7 to 14 atoms per monomer for a total of 1{,}380 electrons, equivalent to the previous largest MP2 level simulation~\cite{Liu2017}. 

Additionally, we apply our program to a crystal structure of an Alzheimer's A$\beta$ 42-residue amyloid fibril (PDB ID: 2BEG). The assembly of amyloid fibrils is a hallmark of Alzheimer's disease, and modeling its formation with force fields presents a significant challenge as this process is predominantly driven by non-covalent interactions \cite{Strodel2021}. Accurate quantum mechanical simulations could therefore elucidate its binding nature, assisting in the development of drugs for Alzheimer's diagnosis and treatment. The 2BEG structure comprises five $\beta$-strands each including 374 atoms. As part of the low latency testing, we employ a variant of 2BEG containing four $\beta$-strands (1{,}496 atoms and 5{,}504 electrons), comprising monomers ranging between 7 and 16 atoms. As the chemical behavior of the above two protein fibril systems are dominated by non-covalent interactions, such structures present ideal biological use cases for MP2. 

All calculations are performed in FP64 precision with the cc-pVDZ primary and cc-pVDZ-RIFIT auxiliary basis sets.

\subsection{Measurement methodology}

The program's correctness has been rigorously tested against existing CPU based implementations in Q-Chem~\cite{qchem4} and GAMESS~\cite{GAMESS}. The analytic gradients have been additionally validated against finite differences for smaller systems and checked for energy conservation.

Program timings are obtained using \verb|MPI_Wtime| at the beginning of each time step in addition to rank local timings of every fragment calculation. 

A FLOP count is performed at runtime by accumulating a local counter on each rank. On every GEMM call with matrix dimensions $(m \times k) \times (k \times n)$ the counter is incremented by $2mnk$. This gives a lower bound on the total number of floating point operations. The counter is accumulated across all ranks with \verb|MPI_Reduce| at the end of the calculation so that an exact lower bound on the floating point performance is output following every calculation.

The code is benchmarked on the time-to-solution (time step latency), scaling, and sustained peak performance. To account for startup effects, the first time step is discarded from the measurements to obtain long-term AIMD throughput.

Where percentages of peak are presented, they are reported with respect to the sustainable peak rather than the theoretical peak to account for thermal throttling.

\section{Performance Results
\label{sec:perf_results}
}

\subsection{Time step latency}

When performing molecular dynamics simulations, there is a critical compromise between size and time scales. Smaller systems allow lower time step latency which gives access to longer simulations times. To demonstrate the capacity of our algorithm to perform large time scale simulations, we perform simulation of the 6PQ5 and 2BEG systems under the micro-canonical (NVE) ensemble on 64 (256 A100 GPUs) and 1{,}024 Perlmutter nodes (4{,}096 A100 GPUs), respectively.

\begin{figure}
    \centering
    \vspace{-0.5cm}
    \includegraphics[width=0.95\linewidth]{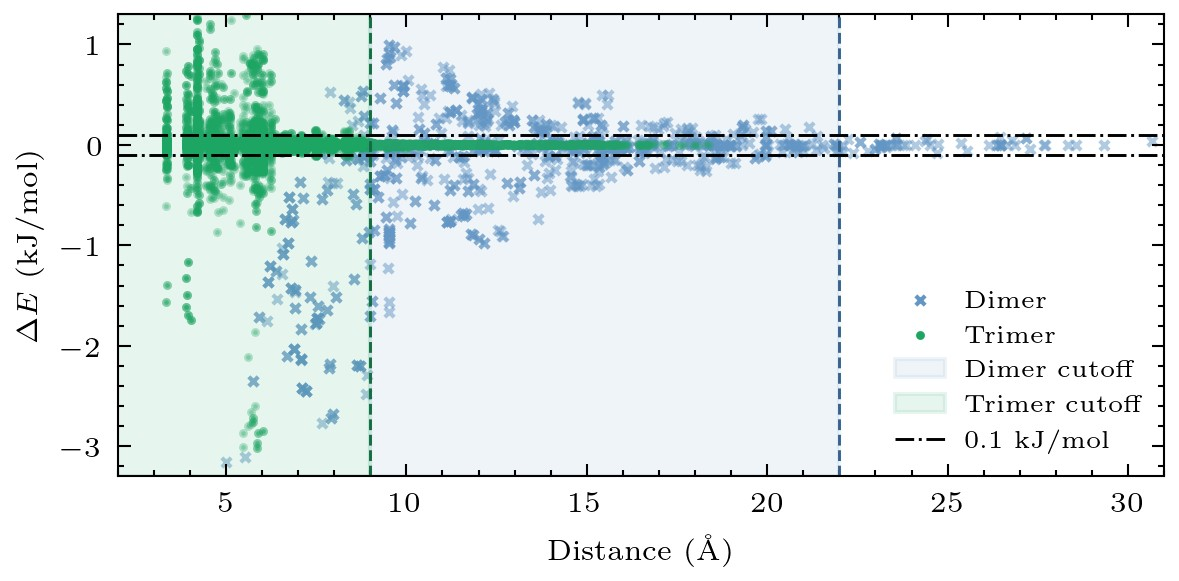}
    \vspace{-0.4cm}
    \caption{Energy contributions of dimers and trimers for the initial geometry of the 6PQ5 system. Cutoff distances used in the calculation are also marked.
}
   \vspace{-0.4cm}
    \label{fig:cutoffs}
\end{figure}

The 6PQ5 system was simulated for 5 ps. To determine the optimal dimer and trimer cutoffs, we evaluated the contributions of the dimers and trimers at the starting geometry. Figure~\ref{fig:cutoffs} shows the diminishing impact of the dimer and trimer corrections as the distance between centroids increases. Contributions become negligible when their magnitude falls below 0.1 kJ/mol, allowing for their exclusion from the calculation. In this case, a dimer cutoff of 22~{\AA} and a trimer cutoff of 9~{\AA} is used, yielding  $\sim$2k total polymers per time step.

\textbf{The 5 ps simulation with 1 fs time steps took 3.16 hours for an average time step latency of 2.27 seconds, enabling simulation of 38 ps/day}. The average floating point throughput was 1.66 PFLOP/s representing 35\% of the sustainable peak. The floating point throughput is primarily limited by the small fragment sizes, which lead to suboptimal GEMM dimensions. In addition, a significant portion of computational time is allocated to FLOP-inefficient $\mathcal{O}(N^3)$ eigenvalue decomposition and integral calculations rather than the $\mathcal{O}(N^4)$ GEMM operations.
With synchronous time steps, the time step latency is increased to 3.0 seconds for the same system, reducing simulation ability to 29 ps/day. Thus, for this system, the asynchronous time step approach provides a 24\% speedup. 

\begin{figure}[t]
\vspace{-0.5cm}
    \centering
    \includegraphics[width=0.95\linewidth]{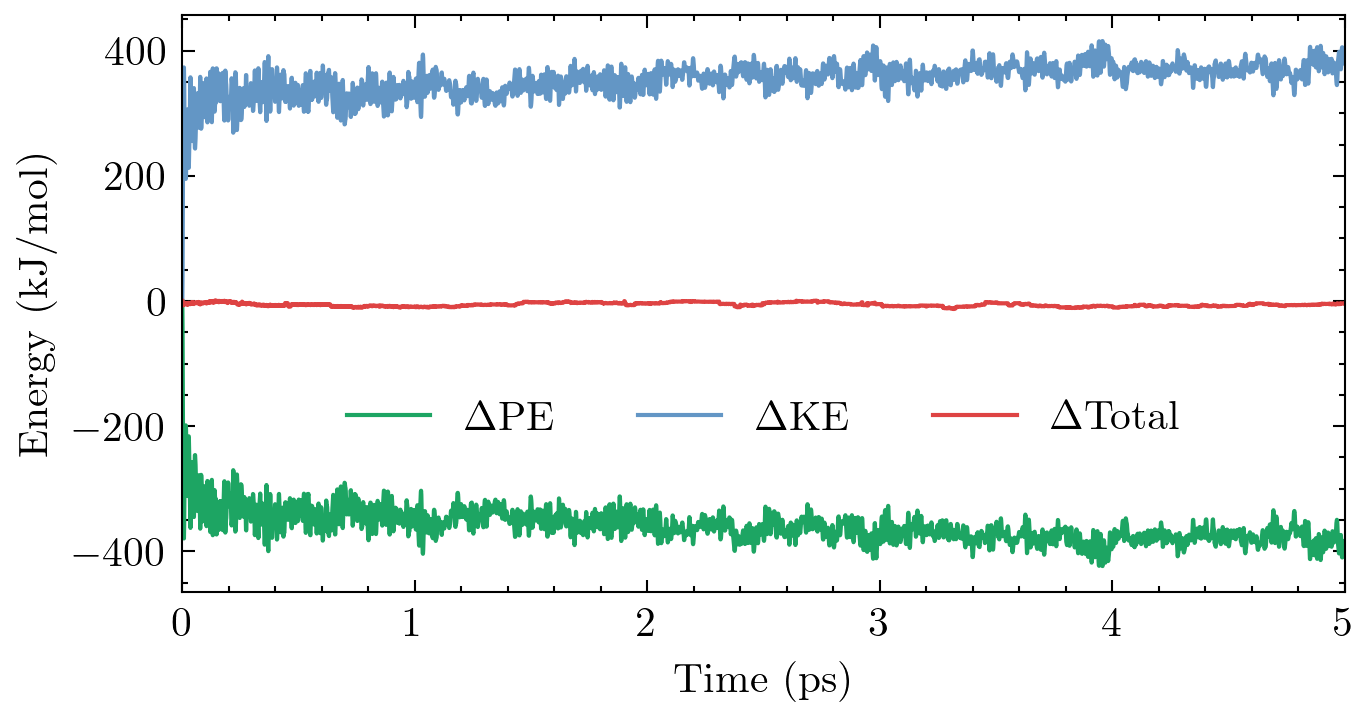}
    \vspace{-0.3cm}
    \caption{Total energy conservation of the AIMD/MP2 simulation for the 6PQ5 system with asynchronous time steps.}
    \label{fig:econsv}
\vspace{-0.5cm}
\end{figure}

A plot of the energy at each time step of the simulation is presented in Figure~\ref{fig:econsv} demonstrating that energy is conserved over the duration of the simulation. This demonstrates the accuracy of the analytic gradients. The small fluctuations in the total energy are due to the time step discretization as well as a small component from polymer corrections dropping in and out as the distance between the polymers fluctuates around the cutoff. It is planned to incorporate a smooth transition for these polymer cutoffs to reduce the effect in future work.

The 4-strand 2BEG system was simulated for 100 fs using 1024 nodes of the Perlmutter supercomputer. The cutoffs were calculated following the same strategy as the 6PQ5 system, using a 20~\AA{} cutoff radius for dimers and 12 \AA{} for the trimers. Each 1~fs step of the simulation took 3.4 seconds. \textbf{This leads to a simulation throughput of 25 ps/day for a system comprised of 1,496 atoms and 5{,}504e$^{-}$ at the MP2 level of theory.} The overall performance of this simulation was 31\% of FP64 R-peak.
With synchronous time steps, the time step latency increases to 5.6 seconds, so the asynchronous time step scheme provides a 40\% increase in throughput by eliminating global synchronization.


\subsection{Scaling}

\begin{figure}[t]
    \centering
     \vspace{-0.5cm}
    \includegraphics[width=0.95\linewidth]
    {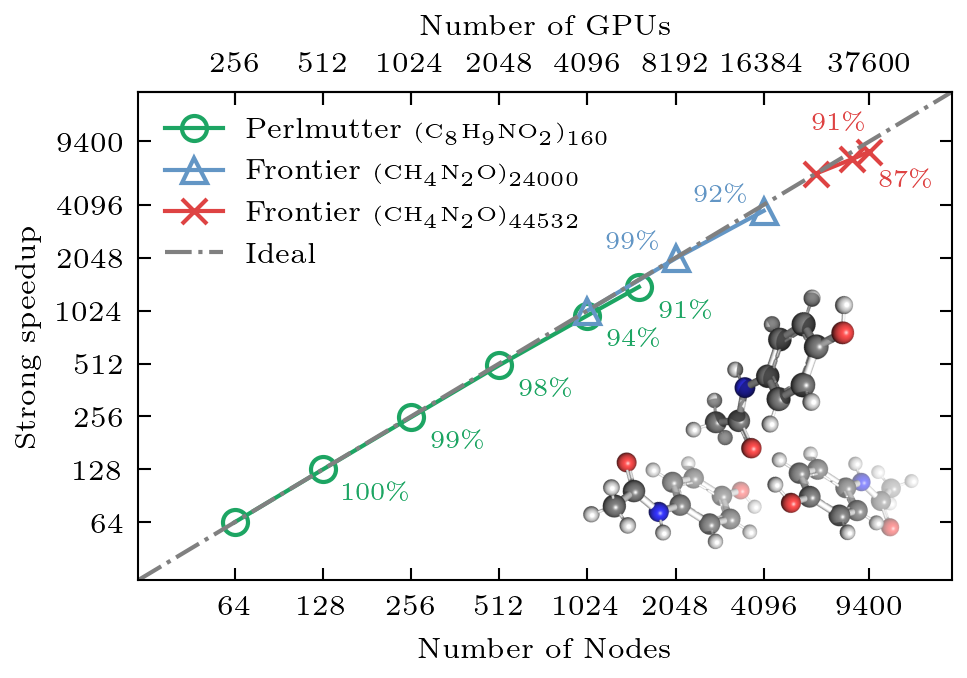}
    \vspace{-0.2cm}
    \caption{Strong scaling of the AIMD implementation from 64 to 1{,}536 nodes on Perlmutter and 1{,}024 to 9{,}400 nodes on Frontier. Points are labelled with the parallel efficiency with respect to the smallest number of nodes.
    }
    \label{fig:strong_1}
\end{figure}

\paragraph{Strong scaling} Figure~\ref{fig:strong_1} shows the strong speedup obtained for a variety of node counts on the Perlmutter and Frontier systems. The program scaled efficiently up to 6{,}144 GPUs (1{,}536 nodes) on Perlmutter (the entirety of the machine), demonstrating 91\% parallel efficiency with respect to the 64-node calculation. The AIMD calculations were performed on a spherical section of a paracetamol crystal lattice with 80 molecules in a dense 36 {\AA} diameter sphere, with one molecule per monomer. On 1{,}536 nodes, an average FLOP rate of 61.31 PFLOP/s was achieved, which corresponds to 54\% of the Perlmutter double precision sustainable peak.

On Frontier, AIMD calculations were performed on two systems built from 24{,}000 and 44{,}532 urea molecules. These systems were fragmented such that there are 4 molecules for a total of 32 atoms and 128 electrons per monomer. Calculations were conducted for 3 time steps on 1{,}024 to 4{,}096 nodes for the 24k urea cluster and 6{,}164 to 9{,}400 nodes for the 44{,}532 urea cluster system. Parallel efficiencies of 92\% and 87\% were were observed for the 24k urea system on 4{,}096 nodes (16{,}384 GPUs, 32{,}768 GCDs) and the 44{,}532 urea cluster system on 9{,}400 nodes, respectively. Additionally, the FLOP rate obtained at each node count was excellent, with 62\%, 61\% and 56\% of attainable peak achieved on 1{,}024, 2{,}048, and 4{,}096 nodes, and of 48\% of attainable peak on 9{,}400 nodes (818 PFLOP/s).

\begin{figure}[t]
    \centering
    \includegraphics[width=0.9\linewidth]{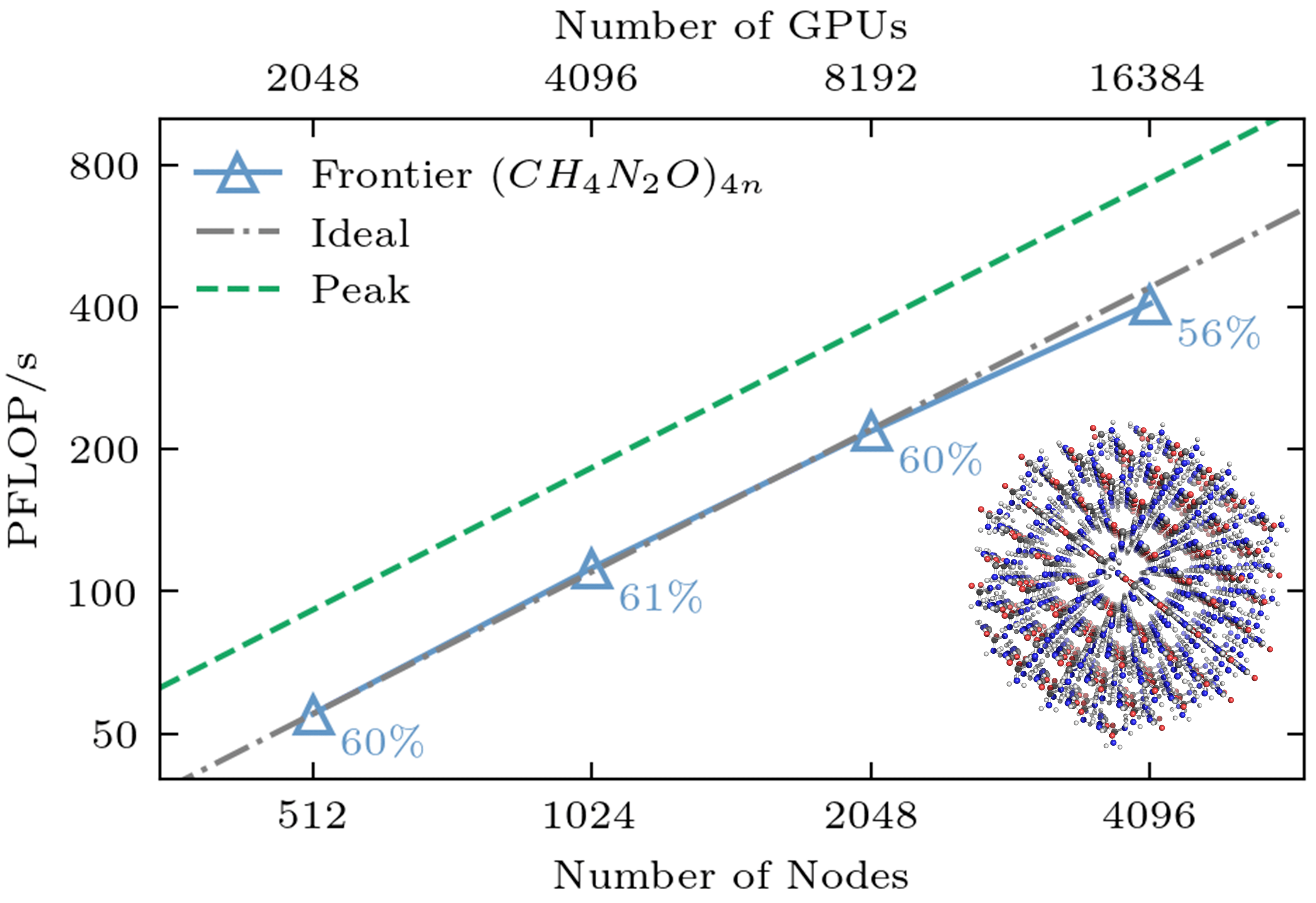}
\vspace{-0.3cm}
    \caption{Weak scaling of spherical urea crystal lattices from 512 to 4{,}096 nodes on Frontier. Points are labelled with percentage of peak obtained.
}
\vspace{-0.3cm}
    \label{fig:weak}
\end{figure}

\paragraph{Weak scaling} 
The weak scaling calculations were performed on systematically growing spherical sections of the urea crystal lattice. The number of fragments in the system was designed to yield an approximately constant workload of 4 polymers per GCD, while progressively increasing the number of GCDs from 4{,}096 to 32{,}768. The resulting weak speedup is shown in Fig.~\ref{fig:weak}. There is a slight drop in efficiency at 4{,}096 nodes which is attributed to increased communication overheads from the dynamic load balancing scheme.


\subsection{Efficiency and peak performance}
\begin{table}[b]
\vspace{-0.5cm}
\centering
\caption{Calculations with record performance and time-step latency.}
\vspace{-0.25cm}
\begin{tabular}{@{}cc@{}}
\toprule
\textbf{Urea}  & \textbf{2BEG}   \\ 
(2{,}043{,}328e$^{-}$)  & (5{,}504e$^{-}$)   \\ 
\midrule
\includegraphics[height=0.17\textwidth]{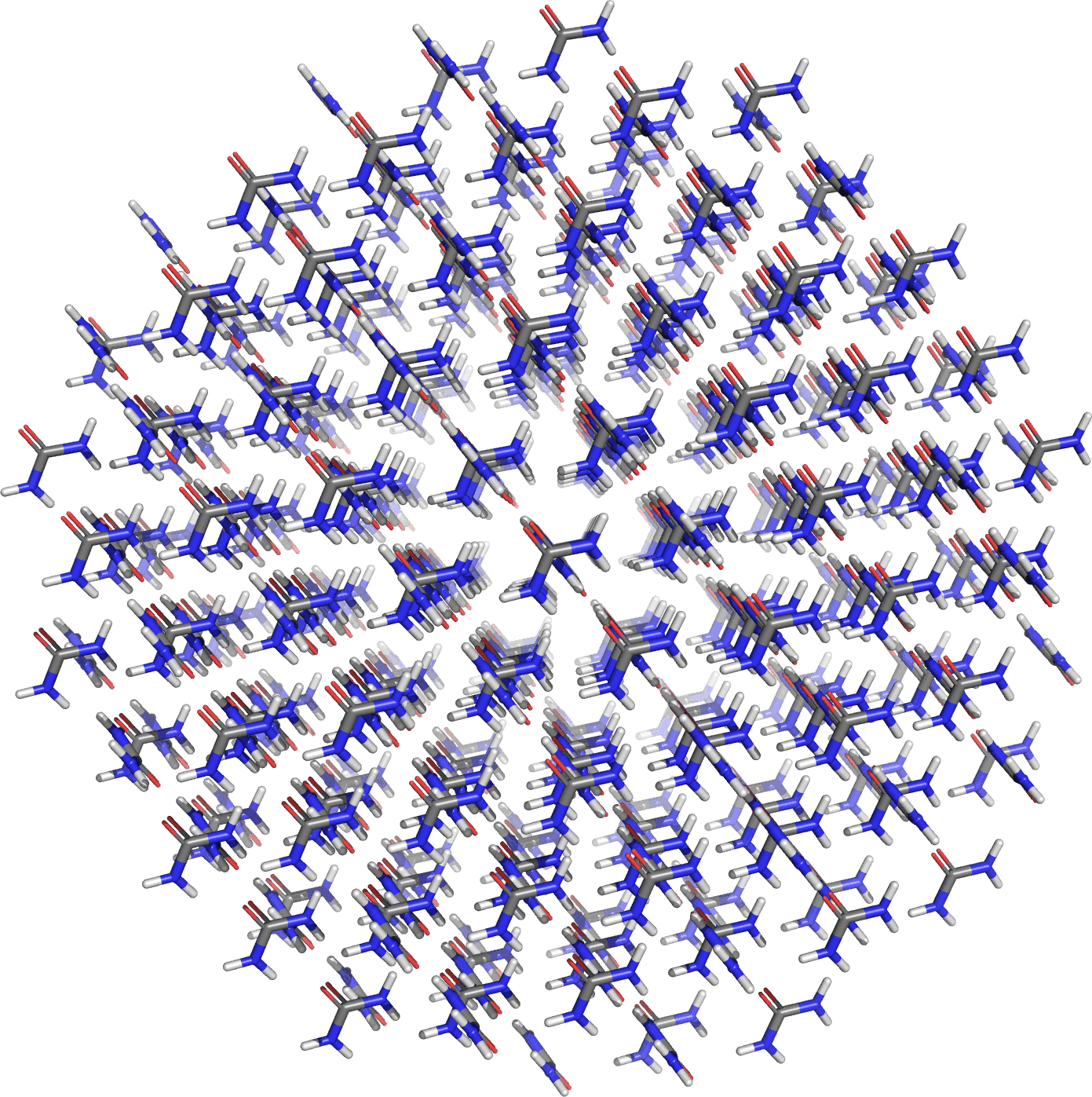}    & 
\includegraphics[height=0.17\textwidth]{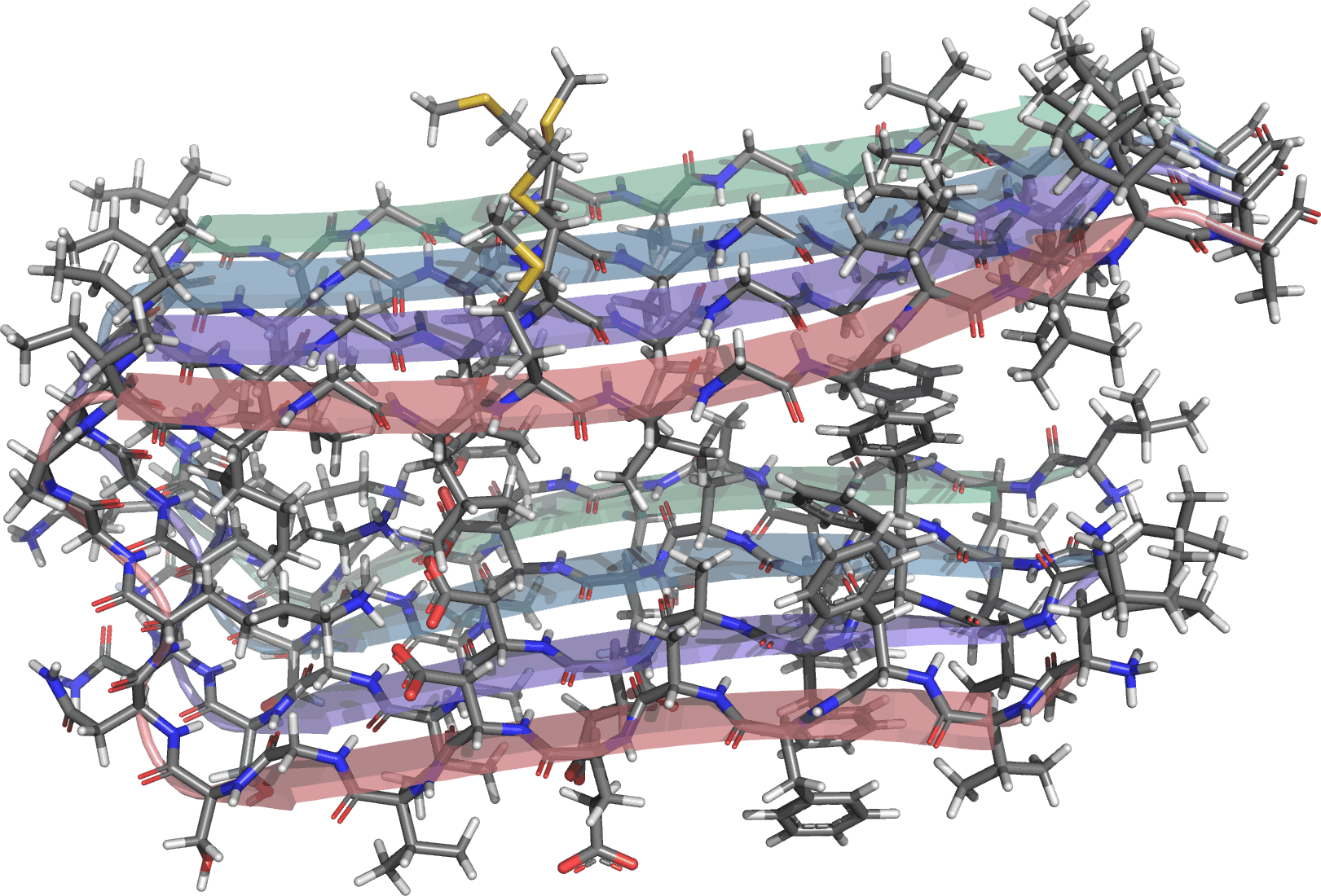} \\
\makecell{Frontier: 9400 nodes \\
1006.7 PFLOP/s} & \makecell{Perlmutter: 1024 nodes\\ 3.4 s/timestep} \\ 
\bottomrule
\end{tabular}
\label{tab:peak_performance}
\vspace{-0.5cm}
\end{table}

To showcase our AIMD program’s ability to
handle large scale simulations on Frontier, we utilized 9{,}400 nodes to perform several steps of AIMD simulations at the MBE3/MP2/cc-pVDZ level on two systems with 44{,}532 and 63{,}854 urea molecules, modeling a total of 1{,}425{,}024 and 2{,}043{,}328 electrons, respectively. Distance cutoffs of 15.3 \AA{} were adopted for dimers and trimers to maintain high-accuracy. For the largest system, this led to $>$2.8 million polymer contributions to the energy and gradient to be evaluated at each iteration. \textbf{Both systems are over 1{,}000$\times$ larger in size than for any previously attempted AIMD at this level of accuracy.}

For the 44{,}532 urea molecule system, the implementation completed a full AIMD time step in 13.7 minutes on 9{,}400 nodes with a floating point throughput of 932.6 PFLOP/s. \textbf{For the 63k urea system with 2.04 million electrons, a full AIMD time step was completed in 25.6 minutes, performing 1.55 zettaFLOPs on 9{,}400 nodes with a FLOP rate of 1006.7 PFLOP/s.
This represents 59\% of Frontier's FP64 R-Peak, marking a milestone in the realm of high-performance computational chemistry for its blend of molecular scale, accuracy, and computational efficiency.}

All calculations presented in this work performed between 31\% and 62\% of the attainable FP64 peak demonstrating efficient usage of the computational hardware. Frontier is one of the most energy efficient machines, ranking 13$^{\text{th}}$ on the June 2024 Green500 list~\cite{Robert2011} with a computational efficiency of 53 GFLOP/joule. Perlmutter ranks 40$^{\text{th}}$ with half the computational efficiency (27 GFLOP/joule). For systems with larger fragments the MI250 based system is far superior, however we observed that for smaller fragments the A100 based Perlmutter significantly outperformed Frontier, likely balancing out the energy efficiency differences with improved hardware utilization. This appears to be due to more efficient random memory accesses improving the integral kernel efficiency and faster vendor provided eigensolver on the A100 system.

\section{Implications}
As discussed in Sections \ref{sec:overview} and \ref{sec:sota}, the effectiveness of  MD methods has historically been limited by two primary hurdles: on the one hand, the lack of accuracy in scalable methods used to calculate forces, on the other, the impractically steep computational scaling with system size of wave-function methods for quantum-accurate \emph{ab initio} forces. These limitations have hindered the applicability of high-fidelity molecular modeling techniques to addressing critical challenges in many areas of chemistry, biology, materials science, and physics, where traditional methods are inadequate due to a simultaneous requirement for high accuracy and large molecular scales.

This work enables a major leap of capabilities in computational chemistry and molecular dynamics, enabling the simulation of biomolecular-scale molecular systems with unprecedented accuracy. We overcome the limitations that previously confined quantum-accurate AIMD simulations to small systems. For the first time, our work allows for the simulation of biomolecular systems containing well over a million electrons, utilizing fully \emph{ab initio} potentials at the MP2 theory level. These simulations are significantly more accurate than any previously conducted at this scale. For systems with $10^3$-$10^4$ electrons, we achieve record time-step latency of a few seconds. This is unprecedented for AIMD calculations using MP2 potential, providing a speedup of $\mathcal{O}(10^3\times)$ for systems with more than $\mathcal{O}(10^2)$ atoms. 

These achievements are made possible through the exploitation of exascale supercomputing resources at  unprecedented computational efficiency in computational chemistry simulations. This is evidenced by attaining 59\% of the Frontier supercomputer's FP64 R-Peak, achieving a 1006.7 PFLOP/s performance on 9,400 nodes. 

This leap forward is not merely incremental; it redefines the boundaries of what is computationally feasible in molecular dynamics, setting a new benchmark for accuracy and efficiency in large-scale simulations. The enhanced scalability and accuracy of our simulation techniques empower the scientific community to tackle longstanding challenges in both chemistry and biology. 

In our study, we have demonstrated through proof-of-concept calculations the potential to tackle two critical problems that have remained unsolved due to their dual requirements for high accuracy and large molecular scale: the identification of crystalline polymorphs and the simulation of the folding and misfolding of amyloid fibrils in Alzheimer’s disease. 
However, our findings pave the way for a wide range of novel applications, including, but not limited to, the exploration of covalent-binding drugs, intricate ligand-receptor interactions dominated by non-covalent forces, detailed simulation of enzyme mechanisms, modeling of biochemical reaction transition states, protein folding processes, DNA interactions with various biomolecules, and an unprecedented level of understanding of the molecular basis of diseases.

Each of these applications not only pushes forward the frontier of scientific knowledge but also lays the groundwork for practical innovations in technology and industry.

\section{Acknowledgments}
This research used resources from the Oak Ridge Leadership Computing Facility (Contract No. DE-AC05-00OR22725) and the National Energy Research Scientific Computing Center (NERSC, ERCAP0026496), supported by the U.S. Department of Energy. RS and EP thank the National Industry PhD program and QDX Technologies for additional support. The authors also thank Dr. Schnoover from Fluid Numerics  for providing access to hardware resources. 

\bibliographystyle{ieeetr}
\bibliography{references} 

\appendix

The following defines the remainder of the intermediates for the RI-HF + RI-MP2 gradient calculation in Eq.~\eqref{eq:rimp2_gradient}. The three-center and two-center gradient coefficients are:
\begin{align}
Z_{\mu\nu}^P &= 4 \Gamma_{\mu\nu}^P + D_{\mu\nu} L_P^{HF} - \frac{1}{2} \sum_{i\sigma} C_{i\mu}D_{\nu\sigma} Y_{i\nu}^P, \nonumber \\ 
\zeta_{PQ} &= -2\sum_{iaR} \Gamma_{ia}^P B_{ia}^R J_{QR}^{-\frac{1}{2}} - \frac{1}{2} L_P L_Q^{HF} + \frac{1}{4}\sum_{ij} Y_{i\mu}^P D_{\mu\nu} Y_{i\nu}^Q,\nonumber
\end{align}
where the auxiliary quantities are defined as:
\[
\begin{aligned}
L_P &= \sum_Q J^{-\frac{1}{2}}_{PQ} \sum_{\lambda\sigma} D_{\lambda\sigma}B_{\lambda\sigma}^Q, \\
L_P^{HF} &= \sum_Q J^{-\frac{1}{2}}_{PQ} \sum_{\lambda\sigma} D^{HF}_{\lambda\sigma}B_{\lambda\sigma}^Q, \\
Y_{i\mu}^P &= \sum_{Q} J^{-\frac{1}{2}}_{PQ} B_{i\mu}^Q, \\
\Gamma_{ia}^P &= \sum_{bjQ}(2X_{ij}^{ab}-X_{ij}^{ba})B_{jb}^Q J^{-\frac{1}{2}}_{PQ},
\end{aligned}
\]
where \(\Gamma_{ia}^P\) is back transformed to the AO basis to form \(\Gamma_{\mu\nu}^P\), and the RI approximated MP2 amplitudes are:
\[
X_{ij}^{ab} = \frac{{(ia|jb)}_{RI}}{\epsij-\epsab}.
\]

The energy-weighted density matrix is:
\[
W_{pq} = \frac{\epspq}{2}D_{pq} \oplus \sum_{qr} G_{ip}^{qr}D^{MP2}_{qr} \oplus L'_{iq} \oplus L''_{aq},
\]
with:
\[
L'_{iq} = \sum_{aP} \Gamma_{ia}^P(P|qa), \quad L''_{aq} = \sum_{iP} \Gamma_{ia}^P(P|iq),
\]
and the anti-symmetrized electron repulsion integrals:
\[
G_{pq}^{rs} = 2(pq|rs)_{RI} - (ps|rq)_{RI}.
\]
Here, \(\oplus\) indicates the sum of intermediate tensors that may not fill the full domain of the resulting tensor.

The density matrix \(D\) is split into HF and MP2 components:
\[
D_{\mu\nu} = D^{HF}_{\mu\nu} + D^{MP2}_{\mu\nu}, \quad D^{HF}_{\mu\nu} = \sum_{i} C_{i\mu} C_{i\nu}.
\]
The MP2 relaxed density matrix \(D^{MP2}\) is defined block-wise in the MO basis:
\[
D^{MP2}_{pq} = D^{MP2}_{ij} \oplus D^{MP2}_{ab} \oplus D^{MP2}_{ai},
\]
where:
\[
D^{MP2}_{ij} = -\sum_{abk}(2X_{ik}^{ab}-X_{ki}^{ab})X_{jk}^{ab},
\]
\[
D^{MP2}_{ab} = \sum_{ijc}(2X_{ij}^{ac}-X_{ij}^{ca})X_{ij}^{bc},
\]
and the occupied-virtual block is obtained by solving the Z-vector equation:
\[
L''_{ai} - L'_{ia} - \sum_{pq} G_{ai}^{pq}\overline{D}^{MP2}_{pq} = \frac{\epsai}{2}D_{ai} + \sum_{bj} G_{ai}^{bj}D^{MP2}_{bj},
\]
where \(\overline{D}^{MP2}_{pq} = D^{MP2}_{ij} \oplus D^{MP2}_{ab}\).

\vspace{12pt}
\end{document}